\documentclass[12pt]{article}

\def\beq{\begin{eqnarray}}
\def\eeq{\end{eqnarray}}
\def\bea{\begin{eqnarray}}
\def\eea{\end{eqnarray}}

\newcommand{\gsim}{\lower.7ex\hbox{$\;\stackrel{\textstyle>}{\sim}\;$}}
\newcommand{\lsim}{\lower.7ex\hbox{$\;\stackrel{\textstyle<}{\sim}\;$}}

\begin{document}

\setlength{\baselineskip}{0.25in}

\begin{titlepage}
\noindent
\begin{flushright}
MIFP-05-35 \\
\end{flushright}
\vspace{1cm}

\begin{center}
  \begin{Large}
    \begin{bf}
A Review of Distributions on the String Landscape \\
%\small{...or How I Learned to Stop Worrying and
%Love the Landscape}

    \end{bf}
  \end{Large}
\end{center}
\vspace{0.2cm}
\begin{center}
\begin{large}
Jason Kumar \\
\end{large}
  \vspace{0.3cm}
  \begin{it}
Department of Physics \\
Texas A\&M University \\
        ~~College Station, TX  77840, USA \\
\vspace{0.1cm}
\end{it}

\end{center}

%\maketitle
\begin{abstract}

We review some basic flux vacua counting techniques and
results, focusing on the distributions of properties over
different regions of the landscape of string vacua and
assessing the phenomenological implications.  The topics we
discuss include: an overview of how moduli are stabilized and
how vacua are counted; the applicability of effective field
theory; the uses of and differences between probabilistic and
statistical analysis (and the relation to the anthropic
principle); the distribution of various parameters on the
landscape, including cosmological constant, gauge group rank,
and SUSY-breaking scale; ``friendly landscapes"; open string
moduli; the (in)finiteness of the number of phenomenologically
viable vacua; etc.  At all points, we attempt to connect this
study to the phenomenology of vacua which are experimentally
viable.

\end{abstract}

\vspace{1cm}

\begin{flushleft}
%hep-th/0601053 \\
January 2006
\end{flushleft}

\end{titlepage}
%\pacs{PACS numbers: }
%]

\setcounter{footnote}{1}
\setcounter{page}{2}
\setcounter{figure}{0}
\setcounter{table}{0}

\tableofcontents

\section{Introduction}

For many years, string theory (or, more generally, M-theory) has
been a promising avenue
for the study of quantum gravity.  In addition, it is a
possible theory which could unify the fundamental forces
experimentally observed in the real world.  One of the most
intriguing aspects of string/M-theory is that it is a theory
with no free parameters.  All low-energy parameters really arise
as fields in string theory, whose values are determined dynamically.
This raises the interesting prospect that perhaps the dynamics
of string theory can uniquely determine the properties of
the real world.

Unfortunately, this hope has slowly given way to the belief that
there are in fact many different string theory solutions, each
of which yields a different low-energy physics.  This set of many
vacua is often called the ``string landscape."  Thus, the question
of how the parameters of the Standard Model are set is simply replaced
by the question of which string vacuum we live in.

In fact, it has seemed clear for quite some time that there are
an infinite number of string theory solutions.  Examples of infinite
classes of vacua which are believed to be stable
include Type IIA string theory in flat space, with
32 supersymmetries and the dilaton as a continuously tunable
parameter.  Other examples include Type IIB string theory in an
$AdS_5 \times S^5$ background with arbitrary integral flux,
and various $N=2$ compactifications on Calabi-Yau manifolds (where
the vacua are parameterized by moduli spaces).

In all of these examples, however, the low-energy physics contains
extended, unbroken supersymmetry.  As a result, these vacua cannot describe
the real world.  It may still be hoped that the constraints
of supersymmetry breaking force us into a unique vacuum.
However, growing evidence suggests that there
are in fact many vacua with no moduli and with weakly broken
supersymmetry\cite{Kachru:2003aw,Acharya:2002kv,flux5,IIA,flux1,flux2,flux3,Giddings:2001yu,
Bousso:2000xa,flux4,Denef:2004dm,Denef:2005mm}
(but see also \cite{kkstringsusybreak,Saltman:2004jh} for evidence
of models with supersymmetry breaking at higher energy scales).

Here we attempt to review some of the arguments in favor and against
the existence of this landscape of vacua.  We also review qualitative
and quantitative statistical results as well as their physical implications
and phenomenological applications.

\subsection{Probabilistic vs. statistical analysis}

Before getting into the details of how these vacua are constructed,
we should first discuss how this information
can be used.  Broadly, there are two types of questions which one
can address by studying the landscape.  A ``what" question asks which
vacuum we live in, but a ``why" question asks why we live in this
vacuum as opposed to any other consistent vacuum.  The distinction
between these two questions is related to the distinction between
probabilistic and statistical analysis of the landscape.

A statistical analysis of the landscape is an attempt to count the
number of vacua which exist in controllable classes and estimate
the distributions of properties among these vacua.  This type of an
analysis can be quite useful in relation to the ``what" question,
because it addresses the point of whether vacua exist whose properties
match the real world to within experimental precision.
If one can show that it
is statistically likely that a vacuum exists whose low-energy properties
are consistent with the real world, then it is likely
that we actually have a vacuum worth looking for\cite{statprob}.
Whether one can actually
find such a vacuum is, of course, a separate challenge which likely requires
a much more sophisticated understanding of the mathematics of string
compactification than we currently have.  But even so, a statistical analysis
of controllable models gives one an idea of where to look for
viable models, or at least a clue as to whether they likely exist.

However, this statistical analysis makes no comment whatsoever
on how nature selects a vacuum.  This
analysis studies only the statistical distribution of properties among
consistent controllable vacua, irrespective of how a vacuum
is chosen.  The goal of this analysis
is to identify the properties of
as many vacua as possible which are phenomenologically viable;  statistics
of this subset are the statistics of a set of vacua, not necessarily
of the real world.
Indeed, we do not know
that nature must choose one of these vacua.  It is simply
an experimental observation that nature may have chosen one of them, given
what we know at the moment.

A probabilistic analysis, however, would seek to determine from this set of
vacua the likelihood that any particular set of low-energy physics is observed.
The idea is essentially to place a probability measure on the landscape, which
determines the likelihood that any particular vacuum is realized.  This measure,
convolved with the statistical distribution of properties on the landscape,
would determine the likelihood that certain low-energy properties are realized by
nature.

One example of a type of probabilistic measure is simply to assert that
every isolated vacuum is equally likely.  Another example would be to
convolve this with an anthropic principle, which states that the only vacua
which are acceptable are those for which it is possible that life can exist.
Thus far, we have made little progress in actually
determining a realistic probability measure (though for some interesting
results, see \cite{selectionprin}).
There does not appear to be a clear approach to determining a selection
principle which is narrow enough to select a unique vacuum.  However,
the statistical analysis of controllable vacua is quite well-defined and is
an area where clear and concrete progress can be made.
As such, we focus mostly on statistical analysis.

This is by no means an attempt at a comprehensive review of work on the
landscape.  Instead, we attempt to survey only one thread of this
field, namely the study of distributions of properties among sets of
(to varying degrees) phenomenologically viable
vacua\cite{grossprop,Arkani-Hamed:2005yv,Saltman:2004jh}.
In section 2, we review the technology used to count flux vacua in
various ensembles.  In section 3, we use this technology to find the
distribution of some low-energy parameters among the set of vacua in
various controlled constructions.  In section 4, we
access phenomenologically interesting questions about string constructions
of the Standard Model with this technology.  We conclude in section 5 with
some interesting questions accessible by this type of analysis.

\subsection{The major caveat -- effective field theory}

Before proceeding, we should discuss one of the major caveats
of this entire analysis ... the reliance on effective field theory.
It is quite difficult to construct consistent string
backgrounds without supersymmetry, and only in very specific
constructions can one find string backgrounds with no moduli.
Instead, the study of the landscape has focussed on an
effective field theory approach.  The basic idea is to look at
10-d string theory compactified on an orientifolded Calabi-Yau 3-fold,
yielding a 4d theory with $N=1$ supersymmetry, expanded about
a flat background.  One
can then derive a low-energy effective field theory ($N=1$ SUGRA)
which describes
scattering in this flat string background.  The crucial jump is to then
look for other solutions to this effective action which are not
asymptotically equivalent to the string background from which the
effective field theory was derived.  Thus, it is perhaps more accurate
to say that we are not in the business of constructing string vacua,
but rather of string-derived effective field theory vacua.

It is an open question whether or not this process is valid
\cite{Banks:2004xh,effFTvalid}.
There is evidence to suggest that string theories formulated
in different backgrounds
are described by different Hamiltonians, not by different states of the
same Hamiltonian.  This might suggest that one cannot begin with a description
of string scattering in one background and bootstrap that into a description
of strings in another background.
One can certainly consider
examples where the potential barrier between the original flat supersymmetric
string solution and the new effective field theory solution is small relative
to the Planck scale.  In such a case, one expects effective field theory to
be valid, at least over a small enough distance scale.  But for a large enough
bubble of false vacuum, the mass of the domain wall is
large enough that it appears to the external observer that a black hole
has formed\cite{Farhi:1986ty}.  The
external observer attempting to find the inflating region
reaches a singularity before reaching an inflating region.  This intuition
would also seem to suggest that we cannot view large bubbles of an
inflating region as excitations of the theory which had
the original flat solution.

However, the fact that de Sitter solutions cannot be reached as excitations
of Minkowski solutions does not necessarily indicate that they do not live
as states in the same theory.  Indeed, a similar result was shown
in \cite{Freivogel:2005qh}, where it was argued that inflating space-times
can never result from an initial pure state, because the inflating space-time
is necessarily described in the dual gauge theory by a mixed state and cannot
arise from a pure-state through unitary
evolution\footnote{Cosmology in the context of AdS/CFT was also discussed
in \cite{Alberghi:1999kd}.}.  But despite the fact that
this mixed state cannot result tunneling or scattering in a pure state, both
the mixed states and pure states are clearly states of the same quantum
Hamiltonian.  Indeed, if there really is a background independent
formulation of quantum gravity, then it seems likely that there should be
a single quantum theory whose Hilbert space describes all possible string
backgrounds.  The expansion about two different fixed backgrounds may be
contained within two subsectors which are disconnected in some limit, but one would
imagine that the full theory should see them both.

And it is certainly the case that, beginning with an effective field
theory derived from a string solution, one can in some cases find new string
backgrounds with different asymptotics.  An example is Type IIB string theory
in $AdS_5 \times S^5$.  One still does not really
know how to quantize Type IIB string theory in the presence of non-trivial
RR-fluxes.  As such, one cannot demonstrate that $AdS_5 \times S^5$ is a consistent
background for the Type IIB string.
Instead one considers Type IIB supergravity, which is the
low-energy effective field theory derived from the scattering of string in
Type IIB string theory in flat space.  One then discovers that this low-energy effective
action has other classical solutions with supersymmetry, including
$AdS_5 \times S^5$.  The consistency of Type IIB string theory in the $AdS$
background has since been given much stronger support by the $AdS/CFT$
correspondence, but the original motivation (confirmed by $AdS/CFT$) came from
effective field theory arguments quite similar to those used in analysis of
the landscape.

Of course, in this case the solution has a large amount of supersymmetry.
However, there seems to be no obvious reason why the arguments against the effective
field theory approach should fail if there is supersymmetry.  Indeed, one of the
arguments against this effective field theory approach is that, according to
$AdS/CFT$, string theory in an $AdS_5 \times S^5$ background with quantized
flux $N_1$ is described by a Hamiltonian which is different from the Hamiltonian
which describes $AdS_5 \times S^5$ with flux $N_2$
(they are both described by $N=4$ $D=4$ SYM, but with different gauge
group rank).  Nevertheless, they all appear as classical supersymmetric
solutions to the equations of motion of the low-energy theory derived from
flat-space.

By no means is this question resolved, and one would certainly hope to develop
a more rigorous demonstration of consistent non-supersymmetric backgrounds which
are phenomenologically viable.  However, given the difficulty in achieving this
aim and the considerations mentioned above, we content ourselves for the
moment with accepting the effective field theory approach at face value.

%*************************************
%*************************************
%*************************************
%*************************************

\section{The Technology of Counting Vacua }

Many different ensembles of vacua have been
studied\cite{Kachru:2003aw,Acharya:2002kv,flux5,IIA,flux1,flux2,flux3,Giddings:2001yu,Bousso:2000xa,flux4,Kachru:2004jr},
but we focus on IIB flux vacua arising from orientifolded Calabi-Yau
compactification, as this is the version on which most work has been done.
At a fundamental level, we are searching for vacua which are phenomenologically
viable, given current experimental data.  As a first step, we search
for 4-d vacua with no moduli and broken supersymmetry.
Of course, generic supersymmetry-breaking
yields potentials for the scalars, lifting all moduli.  However,
without the control of supersymmetry, generic potentials can cause one
or more scalars to run to infinity, destabilizing the solution.  In particular,
one expects that the vacuum energy of a non-supersymmetric solution
contributes a potential term which tends to cause the size moduli to run to
infinity, causing decompactification of the
solution\cite{Dine:1985he,Kachru:2002gs}.

Our aim is then to find supersymmetric vacua in which as many moduli as
possible are fixed,
and then introduce supersymmetry breaking at a scale much lower than
the masses of the scalars.  Because the scalars are fixed in a supersymmetric
vacuum, we have good control over the potential and can be confident that
there are no destabilizing corrections.  Once we introduce supersymmetry
breaking at a lower scale, we can still be confident that the solution is
stable, because the masses of the scalars are much larger than the supersymmetry
breaking scale.  The scalar vevs might shift a little, but they cannot
be destabilized.

It may seem a bit strange to refer to ``supersymmetric" vacua, when a necessary
part of our construction is in fact to break supersymmetry.  To be more concrete,
we imagine an effective potential which we can write in the form
\bea
V= V_{large} +V_{small},
\eea
where there is a controlled limit in which $V_{small} \ll V_{large}$.  We can
then write
\bea
\tilde V =V_{large} +\lambda V_{small}.
\eea
Solutions to the equations of motion are now parameterized by $\lambda$.  We
say that a family of solutions is ``almost
supersymmetric" if the $F$-term
equations\footnote{One might worry that, even if the $F$-terms
vanish, there may be non-trivial $D$-terms which would
break supersymmetry, perhaps badly.  But by the general arguments
of\cite{deAlwis:2005tf,Gates:1983nr,Choi:2004sx}, the
gauge invariance of $N=1$ $D=4$ supergravity implies
that the $D$-terms are of the form
\bea
2 \Re f^{ab} D_b = {\imath k^{ai} D_i W \over W}
\eea
where $f$ is a gauge coupling and $k^{ai}$ generates Killing symmetries
of the K\"ahler metric.  From this, one sees that the $D$-terms
can be non-vanishing only if an $F$-term is
non-vanishing as well.
So we need only check for solutions to the $F$-term
equations.  Note that once we break supersymmetry, we may
still have $|D| \gg |F| >0$; this corresponds to
standard $D$-term breaking in global supersymmetry.}
are solved when $\lambda=0$.  Once we move to $\lambda \neq 0$, the solution
will deform and the
$F$-term equations may no longer be satisfied.  Nevertheless, the corrections to
the solution are small, provided we are in the limit where
$V_{small} \ll V_{large}$.

\subsection{The Type IIB flux vacua ensemble}

With this idea in mind, the plan is to begin with Type IIB string theory
compactified on an orientifolded Calabi-Yau 3-fold.
Much of the quantitative analysis of this set-up was pioneered in
\cite{Giddings:2001yu,Kachru:2003aw,Douglas:2003um,Ashok:2003gk,Denef:2004ze},
and we use their notation.
The CY compactification
preserves 8 real supersymmetries, with the orientifold projection reducing
this number to 4.  The low-energy effective field theory describing this
compactification is $N=1$ $D=4$ supergravity.  This
approximation is valid
in the limit $g_s \ll 1$, $R \gg l_s$, where $R$ is a size modulus for the
compact dimensions.  In the
end, we will consider only those solutions where the moduli are fixed in
this self-consistent regime.
The massless scalar fields are the complex structure moduli, the
K\"ahler moduli and the axio-dilaton.

The potential for this theory is given by
\bea
V=e^K(D_i W \overline{D^i W}^i
-{3\over M_{pl}^2}|W|^2),
\eea
where $K$ is the K\"ahler potential and $W$ is the
superpotential.  Here, $DW$ is a covariant derivative
with respect to the moduli, given by
$DW=\partial W + W\partial K$.
By turning on
NSNS and RR fluxes, we can generate a tree-level superpotential of the
form\cite{Gukov:1999ya}
\bea
W_{tree} =\int G \wedge \Omega,
\eea
where $G$ is the complexified 3-form flux
and $\Omega$ is the holomorphic 3-form.  $G$ depends on the
axio-dilaton and $\Omega$ depends on the complex structure moduli.
These moduli are then fixed by this potential, with masses that
scale as  $m \sim {l_s ^2 \over R^3}$.  In fact,
the $F$-term equations ($DW=0$)
for the complex structure moduli can only be
satisfied when the complex structure moduli assume values such that
the fluxes are imaginary self-dual (ISD).  This means that the
holomorphic structure of the fluxes
must be of the form (2,1) or (0,3).

This leaves only the K\"ahler moduli.  At tree-level, the
no-scale structure of the superpotential implies that the
potential cannot fix the K\"ahler moduli.\footnote{It is worth
discussing this point in a bit of detail.  No-scale
structure arises from a theory with a superpotential which is
independent of the K\"ahler moduli, and with a K\"ahler potential
for the K\"ahler moduli given by $K = -3 \ln[\imath (\rho
-\bar{\rho})]$ (in the case of only one volume modulus).
In this case, any solution of the equations of motion
has $D_{\rho}W \overline{D^{\rho}W}=3|W|^2$, and thus
positive semi-definite potential $V$.
The remaining $F$-term equations are $m$ equations for
$m$ unknowns, where $m$ is the
number of remaining moduli.  If these equations are solved,
then $V=0$ and the K\"ahler moduli
are not fixed (regardless of whether or not the $F$-term equations
for the K\"ahler moduli are satisfied).
Because the superpotential is
independent of the K\"ahler moduli, we also find that supersymmetric
solutions (those which satisfy the $DW=0$ for all moduli)
have $W=0$.
If there are corrections to the K\"ahler potential,
then $V \geq 0$ will no longer be necessary.
But as long as the superpotential is uncorrected, we still
find $W=0$ for supersymmetric solutions.  If the
superpotential receives non-perturbative corrections which
depend on the K\"ahler moduli, however,
then we will generically find
supersymmetric solutions with $DW=0$,
$W \neq 0$, $V \neq 0$.

Note that $W=0$ is not compatible with generic ISD fluxes.
In particular,
non-trivial (0,3) fluxes necessarily force the superpotential
to be non-zero.  (2,1) ISD fluxes are compatible with the
$F$-terms equations and the condition $W=0$.  The breaking of
no-scale structure implies that (0,3) fluxes can also result in
supersymmetric $AdS$ vacua, rather than non-supersymmetric vacua.}
But non-perturbative corrections to
the superpotential and corrections
to the K\"ahler potential\cite{kahlercorr}
(both perturbative and non-perturbative)
break no-scale structure generically.  We will find that
the non-perturbative corrections to the superpotential will
be the dominant correction, and can give masses to the
K\"ahler moduli of order
$\sim {1\over l_s} e^{-{R\over l_s}}$.   These corrections
can arise from gaugino
condensation\footnote{There are in fact two different ways in
which one may model gaugino condensation in $N=1$ SUGRA.  We may
either consider it as an explicit vev for an $F$-term, or as a term
in the superpotential which is exponential in the volume modulus.
The first method is applicable in situations where gaugino
condensation breaks SUSY.
Of course, in such cases the second method will also reveal that,
for appropriate $K$ and $W$, SUSY is broken.
More generically\cite{Nelson:1993nf}, though, the first method can
be used to show that SUSY is unbroken after gaugino condensation.
It is important to keep this distinction in mind, as both
methods appear in the literature.  In this review, we
will represent gaugino condensation via an exponential term in the
superpotential.}
or from D3-instantons,
for example.

We can demonstrate how this can work.
When the complex structure moduli and axio-dilaton are
fixed at tree-level, the superpotential is fixed at
some value $W_0$, which we assume to be negative.
If we also have $|W_0| \ll M_{pl}^2$, then all corrections to
the K\"ahler potential, both perturbative and non-perturbative,
are dominated by the non-perturbative corrections to the
superpotential.  As such, these are the only corrections we
consider.  The corrected superpotential is of the form

\bea
W &=& W_0 + A e^{-a \sigma},
\eea
where $\sigma$ is a real K\"ahler modulus (we assume only one
for the moment, and we set the axion to zero) and
$A$ and $a$ are constants.  The
potential arising from this superpotential fixes the
K\"ahler modulus at finite vev.  Note however, that consistency
of this supergravity approximation requires that
$\sigma \gg 1$.  If we imagine that the correction arises from
a single instanton, then this form is reliable only when
$a\sigma > 1$.  But if $W_0 <0$ and $|W_0| \ll M_{pl}^2$, then
these conditions can be satisfied and our solution is
self-consistent.\footnote{In actuality, the expectation values of
moduli are really determined by extremizing the full potential as
a function of all moduli, not by the procedure described above in
which the light fields are truncated while the heavy fields are
integrated out, with the light fields then reintroduced.  If the
complex structure moduli are integrated out from the full potential,
the effective potential for the remaining K\"ahler moduli will in
fact be slightly different from the one described in the procedure
above\cite{deAlwis:2005tg,deAlwis:2005tf,Choi:2004sx,Yang:2005fa}
due to the backreaction of the heavy fields.
But the essential results of a generic hierarchy of scales
and fixing at finite values of the complex structure and K\"ahler
moduli are unchanged.}

Note that we have not actually shown that the K\"ahler moduli
are fixed\cite{Robbins:2004hx}.  Instead, we
have shown that certain types of
corrections, which can appear in a wide variety of models, will
fix those moduli.  In fact, several specific examples have
been exhibited in which the K\"ahler moduli are explicitly
shown to be fixed by these types of
corrections\cite{Denef:2004dm,Denef:2005mm,IIA}.

It is at this point that one would examine the types of
additional supersymmetry-breaking contributions which
can also be
included\cite{Kachru:2003aw,Brustein:2004xn,Saltman:2004sn}.\footnote{The breaking of supersymmetry
introduces a new subtlety regarding tachyons.  Solving the $F$-term
equations does not necessarily imply that the potential is at a
minimum; some of the masses may in fact be tachyonic.  If the solution
is supersymmetric, however, these tachyons do not constitute an
instability.  But in realistic scenarios in which supersymmetry is
broken and the vacuum energy is positive, one must be sure
to consider only solutions for which the bosonic mass matrix is positive
definite.  It calculated examples this can be true in an ${\cal O}(1)$
fraction of vacua, but in limits (such as near a conifold point) it
can approach zero\cite{Denef:2004ze}.}
An example would be to turn on a small amount of IASD
flux\cite{Saltman:2004sn}.  Another example
would be the introduction of anti-D3-branes.
As argued in \cite{Kachru:2003aw,Giddings:2001yu,Klebanov:2000hb},
the appearance of fluxes leads to a warping of the compactification,
and the potential tends to force the anti-D3-branes to the
warped end of the throat, potentially generating a hierarchy.
One could also rely on non-perturbative
corrections\cite{Brustein:2004xn} to break supersymmetry.

\subsection{Counting the vacua of the IIB ensemble}

Given that we know the form of the potentials,
we can now quantitatively estimate the number of vacua which
we can generate.  By vacua, we mean a choice of fluxes to
turn on and a choice of vevs for the complex structure moduli
and axio-dilaton such that the $F$-term equations are satisfied.

The essential point is simply to determine what inputs
we can vary, and what constraints we are required to satisfy.  Our
inputs are the choice of complex structure moduli and the
choice of integer-valued fluxes.  Our constraints are the $F$-term
equations and the cancelation of RR-tadpoles.  In the simplest
cases, the only non-trivial RR-tadpole condition arises from the
cancelation of space-filing D3-brane charge, and can be written as
\bea
L_* = N_{D3} + {1\over 2} \int F_{RR} \wedge H_{NSNS},
\eea
where $-L_*$ is the D3-brane charge carried by the orientifold,
$N_{D3}$ is the number of D3-branes we have added in by hand,
and $L={1\over 2} \int F_{RR} \wedge H_{NSNS}$ is the quantized
amount of D3-brane charge induced by turning on the RR and NSNS
3-form fluxes.  Given our GVW superpotential, the tree-level
equations of motion are satisifed only if our fluxes are
imaginary self-dual (ISD)\cite{Giddings:2001yu}, which in turn
is sufficient to
show that $L>0$.  Supersymmetry then requires that $N_{D3}>0$.

For convenience, we absorb the
$e^K$ factor in front of the potential into the rescaling
$DW(z) \rightarrow e^{K\over 2} DW(z)$, $\Omega(z) \rightarrow
e^{K\over 2} \Omega (z)$.  As a result, we drop the overall
factor of $e^K$ in front of the potential, but instead find
that $DW$ and $\Omega$ are no longer holomorphic functions of
the moduli.
We denote by $n=h^{2,1}_{-}$ the number of complex structure moduli
which survive the orientifold projection.  The number of real fluxes
which we can turn on is then given by $4n+4$.  We can represent
the choice of flux in an integral basis by the
vector $\overrightarrow{N}$.  The charge
$L$ induced by the flux is a form which is quadratic
in $\overrightarrow{N}$.  For ease of calculation, we will also
end up ignoring the quantization of the flux vector
$\overrightarrow{N}$, allowing us to replace sums over fluxes
with integrals.  This approximation is justified\cite{Denef:2004ze}
in the limit where $L_* > n$.

Our strategy\footnote{This can be related to the attractor
mechanism\cite{Denef:2004ze,attractor}.} for
counting vacua is to integrate the unit element
over all
choices of complex structure moduli and sum over all possible
choices of flux, while using a $\delta$-function to impose
the $F$-term equations and using a step function to impose
tadpole cancelation.
The number of flux vacua ${\cal N}$ can thus
be expressed as

\bea
{\cal N} &=& \sum_{fluxes} \int d^{2n+2} z \,
\delta_z ^{2n+2}(D_z W) \, \theta (L_* -L) \nonumber\\
&=& \int {d\alpha \over \alpha} e^{\alpha L_*}\,
\int d^{4n+4}N \, \int d^{2n+2}z e^{-\alpha L} \,
\delta_{DW} ^{2n+2}(DW) \det|D^2 W|.
\eea
Since $L$ is quadratic in $N$, we may rescale $N \rightarrow
{N\over \sqrt{\alpha}}$ to get
\bea
{\cal N} &=& \int {d\alpha \over \alpha} e^{\alpha L_*}
\alpha^{-2n-2}\, \int d^{4n+4}N \, \int d^{2n+2}z e^{-L} \,
\delta_{DW} ^{2n+2}(DW) \det|D^2 W| \nonumber\\
&=& {(L_*)^{2n+2} \over (2n+2)!} \,
\int d^{4n+4}N \, \int d^{2n+2}z e^{-L} \,
\delta_{DW} ^{2n+2}(DW) \det|D^2 W|.
\eea
At this point, we must find a nicer basis
for the fluxes.  Since the fluxes appear in the superpotential, we
can write a basis for the fluxes in terms of the superpotential
and derivatives thereof (with respect to the complex structure
moduli).  This is particularly useful for integrating over the
$\delta$-function in a simple manner.

We choose a basis for the fluxes where the coefficients are
given by
\bea
X&=& W \nonumber\\
Y_A &=& D_AW \nonumber\\
Z_I &=& D_0 D_I W,
\eea
where $A=1...n$, $I=0...n$.  $A$ is an index which runs over
the complex structure moduli, while $I$ runs over those moduli
plus the axio-dilaton ('$0$') as well.
In terms of these coefficients, we
may write the D3-brane charge induces by the fluxes as
\bea
L =|X|^2 -|Y|^2 +|Z|^2.
\eea
We get the expression
\bea
{\cal N} &=& {(L_*)^{2n+2} \over (2n+2)!} \,
\int d^{2n+2}z\, \int dX \, d^{2n+2}Z \,
e^{-|X^2|-|Z^2|} \,
\det|D^2 W|_{Y=0}.
\eea
The $\det |D^2W|$ factor arises from the change in
normalization of the $\delta$-
function which allows it to fix a flux instead of a modulus.
This determinant depends on the fluxes $X$ and $Z$, the complex
structure moduli and axio-dilaton, and on the geometric
data ${\cal F}$ of the
Calabi-Yau.  Note the way in which the number of vacua
factorizes.  The prefactor depends on only on very simple
topological data ($n$ is the number of complex structure
moduli and $L_*$ is the background charge of the orientifold,
which is the same as the Euler character of the Calabi-Yau
four-fold which is the $F$-theoretic dual of this compactification).
This prefactor essentially counts the number of choices of the
flux which are compatible with the RR-tadpole conditions at any
given point in moduli space\footnote{Thus far, we have basically
ignored the discreteness of the flux.  If $n \gg L_*$, many
integral flux coefficients vanish and the discreteness cannot
be ignored.  In this case, one instead expects ${\cal N} \sim
e^{\sqrt{2\pi c n L_*}}$, where $c$ is a
constant\cite{Ashok:2003gk}.}.  The integral essentially integrates
over a density of vacua in complex structure moduli space.

An index density can be defined
by removing the absolute value symbols from
the determinant, and counts the vacua weighted by a sign.  This
index density provides a minimum
estimate for the number vacua, and one
of the reasons for its usefulness is the simple estimate
demonstrated in\cite{Ashok:2003gk,Denef:2004ze}:
\bea
\rho_{ind} &=& \pi^{-(n+1)}\det (R+1\omega)
\eea
where $R$ and $\omega$ are the curvature 2-form and K\"ahler 2-form
respectively on the complex structure moduli space ${\cal M}_C$.  This
indicates that string vacua tend to be concentrated in regions
of the moduli space of high curvature, such as near a conifold singularity.
The correlation between the density of vacua and the curvature of the
moduli space is key to developing a correlation between low-energy
parameters and the density of vacua.\footnote{There are
topological arguments to suggest that this type of
index density estimate should be valid beyond the limited scope of
Calabi-Yau compactification\cite{Ashok:2003gk}.}

We have not yet included the non-perturbative corrections
to the superpotential.  These generate additional
$F$-term equations which fix the remaining K\"ahler moduli.
The form of this equation depends on the
non-perturbative corrections, but does not seem to
depend on the parameters relevant for our original counting (i.e.,
$n$, $L_*$, ${\cal F}$).  Indeed, it seems to depend on the
tree-level superpotential only through $W_0$.  As such,
this problem essentially factorizes into a product of solutions to
the complex structure moduli (and axio-dilaton) $F$-term equations
and K\"ahler moduli $F$-term equations.  This counting of
solutions to the K\"ahler moduli equations of motion
depends on the exact form of the non-perturbative corrections, so
we do not address it in detail.  But we shall see that we can gain
much statistical knowledge simply from the distribution of solutions
of the first set of $F$-term equations.

Finally, remember that we still have some self-consistency requirements,
namely that the moduli be fixed in such a way that $g \ll 1$ and
$R \gg 1$ (equivalently, $\sigma \gg 1$).  This analysis is only
valid for the fraction of
vacua where the moduli happen to be fixed in this self-consistent
region.  But the distribution of vacua is basically uniform in this
region, so restricting to this region does not dramatically reduce
the number of vacua\cite{Ashok:2003gk}.

\subsection{Cosmological constant}

We may similarly study the distribution of the tree-level
cosmological constant by counting the flux vacua for which
cosmological constant is within a specified range.  For a
supersymmetric
solution where the $F$-term vanishes,
the cosmological constant is simply given by

\bea
\Lambda =-3e^{K}{|W|^2\over M_{pl}^2}.
\eea
What really are seeking is the distribution of
$e^K |W|^2$, which can be written very simply in terms of
our fluxes.  Indeed, this term is precisely the flux
coefficient $X$ which we defined earlier.

We can now count the number of flux vacua whose
tree-level cosmological constant is given by $|\Lambda|
\leq \Lambda_0$.  We simply follow the same counting
procedure as before, except we include an integration
over a parameter $\Lambda$, and a $\delta$-function in our integral
to pick out vacua with cosmological constant equal to $\Lambda$.
This yields
\bea
{\cal N}&=& \int {d\alpha \over \alpha} e^{\alpha L_*}\,
\int_0 ^{\Lambda_0} d\Lambda
\int d^{4n+4}N \, \int d^{2n+2}z \, e^{-\alpha L} \,
\delta_{DW} ^{2n+2}(DW) \delta(\Lambda -|W|^2)\det|D^2 W|.
\nonumber\\
\eea
We can again rescale $N \rightarrow {N\over \sqrt{\alpha}}$,
remembering now to rescale $\Lambda$ and
$\Lambda_0$ by ${1\over \alpha}$
as well.  We then find
\bea
{\cal N}&=& \int {d\alpha \over \alpha} e^{\alpha L_*}
\alpha^{-2n-2}\,
\int_0 ^{\Lambda_0 \alpha} d\Lambda
\int d^{4n+4}N \, \int d^{2n+2}z \, e^{-L} \,
\delta_{DW} ^{2n+2}(DW) \delta(\Lambda -|W|^2)\det|D^2 W|.
\nonumber\\
\eea
The vacuum density on moduli space is now a function of the
parameter $\Lambda_0$ as well.  If $\Lambda_0$ is small enough,
then the
vacuum density is approximately constant with respect to $\Lambda$
in the range $0 \leq |\Lambda| \leq |\Lambda_0 |$
and is equal to it its value at $\Lambda =0$.  We get
\bea
{\cal N}&=& \int {d\alpha \over \alpha} e^{\alpha L_*}
\alpha^{-2n-1} \Lambda_0 \,
\int d^{4n+4}N \, \int d^{2n+2}z e^{-L} \,
\delta_{DW} ^{2n+2}(DW) \delta(|W|^2)\det|D^2 W|
\nonumber\\
&=&\Lambda_0 {(L_*)^{2n+1} \over (2n+1)!}  \,
\int d^{4n+4}N \, \int d^{2n+2}z e^{-L} \,
\delta_{DW} ^{2n+2}(DW) \delta(|W|^2)\det|D^2 W|.
\eea
Making the change of flux basis as before, and integrating
over the flux coefficient $X$ as well gives us
\bea
{\cal N} &=&\Lambda_0 {(L_*)^{2n+1} \over (2n+1)!}  \,
\int d^{2n+2}Z \, \int d^{2n+2}z e^{-|Z|^2} \,
\delta_{DW} ^{2n+2}(DW) \det|D^2 W|_{X=Y=0}.
\eea
This expression for the number of flux vacua is similar to
the expression we found earlier, but with two notable differences.
The vacuum density on complex structure moduli space is slightly
different.  But more importantly, the  previous scaling of
$L_* ^{2n+2}$ is now replaced by a scaling of $L_* ^{2n+1} \Lambda_0$.
In particular, the number of flux vacua with tree-level cosmological
constant between $\Lambda_0$ and $0$ scales linearly
with $\Lambda_0$, when $\Lambda_0$ is small.

\subsection{Open string moduli}

So far, we have only discussed the fixing of closed string
moduli.  If there are space-filling branes in the theory, then the low
energy effective field theory also has scalar fields arising from
open strings.  These scalars are in some ways less problematic than
closed string scalars.
It is important that all of these scalars get a mass for phenomenological
reasons, and supersymmetry breaking will generically allow potentials
which can give mass to all of these scalars.  We wanted to make sure that
the closed string scalars got mass at an even higher scale in order to ensure
that the generic potentials do not ruin the compactification.
The endpoint of open string
scalar decay is usually much more controllable, however, and can be understood
in terms of brane decay.  In any case, we do not need to worry about the
runaway problems which plague closed string moduli, since the open string
scalars are compact.
As a result, we can rely on supersymmetry breaking
effects, if needed, to generate open string masses.  Indeed,
these effects are needed to fix
moduli in realistic cases.

We can broadly divided up open string scalars into two groups: those
which have vector-like gauge transformations,
and those which do not.  If a chiral multiplet
has vector-like transformations,
then the entire multiplet can pair up with another multiplet
and gain mass.  But if the multiplet's gauge transformations under the low-energy
gauge group are not vector-like, then the fermionic component cannot get mass.
This implies that its scalar superpartner also cannot
get mass unless supersymmetry is
broken\footnote{In fact, the scalars may have an effective
mass at the cosmological constant scale, even if supersymmetry
is unbroken.  As this scale is already low enough to be ruled
out by experiment, the larger supersymmetry
breaking scale still controls the mass of these scalars in
realistic models.}.
Since the Standard Model has
fermions which are chiral under the Standard Model gauge group, the
associated squarks and sleptons can only get masses set at the supersymmetry
breaking scale.

More generally, non-trivial dependence of the superpotential on the open
string scalars may help avoid certain difficulties in realizing inflation
on the landscape\cite{Kachru:2003sx}.  It was shown in \cite{Kachru:2003sx}
that models of
inflation where the inflaton potential arose from brane-antibrane interactions
in a warped geometry could satisfy the slow-roll conditions.  However, this
condition would be spoiled by the simple mechanism for fixing the K\"ahler
moduli described above (a non-perturbative correction to the superpotential
which introduces dependence on the K\"ahler modulus $\rho$).  This problem
may be alleviated, however, by a fine-tuned dependence of the superpotential
on the open-string scalar which acts as the inflaton.

There are a variety of effects which can give rise to masses for open string
scalars, both SUSY-breaking and SUSY-preserving.  For example, ISD fluxes can give
SUSY-preserving masses to vector-like chiral multiplets living on D7-branes (though
not D3-branes).  IASD fluxes give soft supersymmetry-breaking
masses to some open string
scalars.\footnote{(0,3) ISD fluxes would also
give soft
SUSY-breaking masses to open string scalars
if no-scale structure were unbroken.  But as we reviewed earlier, one expects
that generic corrections to the superpotential and K\"ahler potential will
break no-scale structure.}  Also, the open string scalars will appear
in the D-term potential along with FI-terms, which depend on the closed
string K\"ahler moduli.  But a generic type of supersymmetry breaking added to
any supersymmetric compactification will also generate a generic potential for
these scalars whose scale is set by the SUSY-breaking scale.

\subsection{Other ensembles}

One may in a similar way construct flux vacua in other string
settings.  Each of these settings provides a different statistical
distributions.    Recently, significant work has been done on
the study of flux vacua of Type IIA compactified on an orientifolded
Calabi-Yau
three-fold\cite{Kachru:2004jr,flux5,IIA}\footnote{Although this
ensemble might appear to be simply the mirror of
our IIB ensemble, it is in fact somewhat different.  The difference
arises from the fluxes of IIB and IIA, which do not map into
each other in a simple way under mirror symmetry.  Instead, the
fluxes map into geometric fluxes, and may result in mirrors
which are non-K\"ahler, and perhaps even
non-geometric\cite{flux1,nonkahlergeom}.  We
will not discuss this further, except to note that
when all such additional compactifications are included in the
ensemble, then we expect the distributions of properties in the
extended IIA ensemble to be identical to that of the mirror
extended IIB ensemble.}.  It appears to be
much easier to find explicitly controlled solutions in
these examples, but with statistical distributions which
are much less broad.  We will not describe these constructions
or the relevant statistical analysis in detail; we
simply note the important results and refer the reader to the
original sources for more information.

In the ensemble of Type IIA flux vacua on orientifold CY's, we
are at liberty to turn on the RR-form fluxes $F_0$, $F_2$ and $F_4$
as well as the NSNS-form $H_3$.  The potential which fixes
the complex structure moduli depends only on the various RR-fluxes
($F_{0,2,4}$) while the K\"ahler moduli are fixed by the NSNS-flux
$H_3$.  The choice of $F_0$ and $H_3$ is
fixed by the constraints of tadpole cancelation, but the
choice of $F_4$ is
arbitrary.  This choice gives us an infinite number of
flux vacua.  However, as $|F_4| \rightarrow \infty$, these vacua will
concentrate at the limit of small
coupling and large volume.  As a result,
there are an infinite number of controllable vacua, but not
phenomenologically viable vacua (for large enough volume of the compact
space, these vacua can be ruled out by experiment).

In this construction, one finds that the number of vacua in which
the size of the compact manifold is $\leq R_*$ scales as
${\cal N} (R \leq R_*) \sim (R_*)^4$.  Similarly, the number of vacua
with small cosmological constant scales as
${\cal N}(|\Lambda| \leq |\Lambda_*|) \sim
(|\Lambda_*|)^{-{2\over 9}}$.  The vacua of this ensemble
are most highly populated at small coupling, small cosmological
constant and large compact volume.

Another ensemble subject to recent interest is the set of flux
vacua arising from a compactification of $M$-theory on a $G_2$
manifold with 4-form fluxes
turned on\cite{Acharya:2002kv,Acharya:2005ez}.

%**************************
%**************************
%**************************
%**************************

\section{Distributions}

Having seen how to develop the technology for counting
flux vacua in a variety of string compactifications, we
would like to use this technology to study the distribution
of low-energy observables over the set of vacua.  The main
points we need to understand are how low-energy
parameters depend on the fluxes and moduli, and how the
density of vacua correlate with the choices of flux and moduli.

Note that this is not a survey of the distribution of low-energy
parameters over all vacua.  It is only a survey of the
distribution over certain controllable sets of vacua.  These
distributions may be representative of the distribution over
all vacua, but there is not yet sufficient evidence to suggest
that this possibility is correct or incorrect.  We can only say
so far that we are studying the distribution of parameters in a
set of vacua which model-builders can potentially construct.

\subsection{Gauge group rank}

As an example, one can compute the distribution of gauge
group rank arising from D3-branes in the IIB ensemble
using the basic
counting methodology we have already
reviewed\cite{gaugegrouprank}.  The basic
point is that the number of flux vacua which
we have found is given by
\bea
{\cal N} =c_{CY} L_* ^{2n+2},
\eea
where $c_{CY}$ is a constant which depends on the structure of
the Calabi-Yau,  $L_*$ is the D3-brane charge of the orientifold plane
and $n$ is the number of complex structure moduli.  From our
previous derivation of this result, one sees that the $L_*$
dependence arises entirely from the step function which was
inserted to enforce the condition that the charge arising from
the fluxes be bounded by $L_*$.  Thus, by an appropriate choice
of the step function, we find that the number of flux vacua with
flux charge less than any integer $L$ is given by ${\cal N} =
c_{CY} L^{2n+2}$.  Since the number of D3-branes which must be
added by hand (to cancel the RR tadpoles) is simply given by
$N_{D3}=L_*-L$, we can easily compute the average number of D3-branes
in this ensemble of vacua:

\bea
\langle N_{D3} \rangle &=& L_*-{1\over c_{CY}L_* ^{2n+2}}
\int_0 ^{L_*} dL \, L\partial_L (c_{CY} L^{2n+2})\nonumber\\
&=& {L_* \over 2n+3}.
\eea
Here we see that the ensemble average is entirely independent of
the details of the Calabi-Yau, including
its singularity structure, geometric
data, etc.  Instead, it depends entirely on topological properties
of the compactification.

Using this average, one may characterize the entire distribution.  In
particular, the fractional density of vacua as a function of the rank
$R$ of the gauge group is given by
\bea
\rho &=& -\partial_R ({L_* -R \over L_*})^{2n+2} \nonumber\\
&\sim& -\partial_R (e^{-{R\over \langle N_{D3} \rangle}}) \nonumber\\
&\sim& {1\over \langle N_{D3} \rangle} e^{-{R\over \langle
N_{D3} \rangle }}.
\eea

One can make a similar calculation in the limit where the tree-level
cosmological constant is small, and one finds that the fractional rank
distribution has the same form of exponential dropoff, but instead
characterized by a modified average
\bea
\langle N_{D3} \rangle_{small c.c.} = {L_* \over 2n+2}.
\eea
From this we see that there is indeed a small correlation between
the rank of the gauge group (arising from D3-branes) and the tree-level
cosmological constant.  This correlation goes as ${1\over n}$, and thus
becomes small in the limit where the number of complex structure moduli
is much larger than the $L_*$, the background D3-charge of the orientifold
planes.  This type of correlation between different low-energy observables
on the landscape (in this case, between cosmological constant and gauge
group) is very interesting from a phenomenological point of view.  Strong
correlations, combined with experimental input, can provide a very useful
guide for string model-building.

Note also that we have only considered the fixing of closed string
moduli.  To truly count the number of vacua, we should convolve this
distribution with the density of solutions to the equations of motion
for the open string moduli as well.
Douglas\cite{Douglas:2003um} has argued
that the number of solutions to the equations of motion for open string
moduli should go as $c^R$, where $c$ is a constant and $R$ is the rank
of the open string gauge theory.  A more detailed study of the distribution
of vevs for open string moduli is presented in \cite{Gomis:2005wc}.

The choice of which distribution one uses depends on the type
of question one asks.  For example, suppose a model-builder was attempting
to decide if a particular choice of Calabi-Yau manifold would be a good
choice for study, based on the likelihood that it contains a model with
suitable hidden sector gauge group at high energy.
In this case, one would first worry
about the the closed string dynamics which leads to a choice of gauge
group, and only later would concern oneself with the distribution of open
string vacua.  On the other hand, if one is studying the total number of
vacua, it would make sense to study the entire distribution of solutions for
open and closed string scalars.

\subsection{Fluxes and moduli}

More generally, we would like to understand how our choice of
integral flux couples to our possible choice of moduli through
the $F$-term equations.  This determines how broadly low-energy
parameters are distributed.
A useful parameter for estimating our ability to fix moduli
is\cite{IIA}

\bea
\eta = {\rm \# \, of\, real\, participating\, fluxes
\over \# \, of\,  real\,  moduli},
\eea
where the ``participating" fluxes are those which enter the
terms of the potential which involve the set of moduli
in question.  This ratio is useful because we may think
of the $F$-term equations for $q$ moduli as really being $q$
equations for the participating fluxes, with the moduli being
parameters.  The more participating fluxes there are (i.e., higher
$\eta$), the easier it is to solve the $F$-term equations for
arbitrary values of the moduli.  The smaller the number of
participating fluxes (smaller $\eta$), the more we must fine-tune
the moduli to allow $F$-term equation solutions, if they can be
found at all.  Thus, for $\eta >1$ we expect to be able to fix
the moduli with a broad distribution of possible solutions by an
appropriate choice of flux.  For $\eta \sim 1$, we still expect
that we can probably fix almost all moduli, but the discreteness
of our choice of flux might limit the range of moduli values we
can achieve by tuning fluxes.
For $\eta <1$, however, we expect
that we only have enough freedom to fix a fraction of the moduli.

Our original IIB ensemble yields
$\eta_{IIB,cpx} \sim 2$, when we look only at the complex
structure moduli.  This indicates that we can easily fix the
complex structure moduli over a large range of the complex
structure moduli space.  However, the fluxes do not let us
fix the K\"ahler moduli at all; other non-perturbative effects
are required for this.

In the Type IIA case, we find that for the complex structure moduli
and K\"ahler moduli we have $\eta_{IIA,cpx} \sim {1\over 2}$ and
$\eta_{IIA,k} \sim 1$ respectively.  This tells us that we can
fix all the K\"ahler moduli, but not along a particularly broad
distribution.  Only half of the complex structure moduli are fixed,
however.  The remaining half are unfixed axions.
For each such axion,
there is a Euclidean D2-instanton which can give mass to the
axion.
In any case, since the axions are compact scalars, we again
will not need to worry about the standard
runaway problems of closed string moduli and can thus also rely on
supersymmetry-breaking corrections to generate masses for the
axions.  Note that experimental constraints on the masses
of possible axions are looser than those on non-axionic scalars.

In the ensemble of flux vacua obtained by studying $M$-theory
compactified on a $G_2$ manifold\cite{Acharya:2005ez}, we
similarly find $\eta_M \sim {1\over 2}$, again implying that
we can only fix half of the moduli.

\subsection{``Friendly landscapes" and the scanning of parameters}

We have found that in the IIB ensemble, certain observables
such as $W$ (which is related to the cosmological constant and
the gravitino mass) have very broad distributions on the
landscape.  This is due to the high value of $\eta$ in this
ensemble; there are enough fluxes so that for a broad distribution
of possible values of $W$, one can find a choice of fluxes and
moduli which satisfy the $F$-term equations.

But this is based on a hidden assumption, namely
that the GVW superpotential generates somewhat generic
couplings for the moduli.
If all the moduli couple to each other in the potential, then
we can expect our intuition to hold.  This manifests itself in
our formalism as a condition on the matrix $\det |D^2 W|$, which
depends on our flux coefficients $X$, $Y_A$ and $Z_I$ (in this
basis, for example, $X$=$W$), the
complex structure moduli and on the geometric
data of the Calabi-Yau manifold.  The geometric data  encode
for us the way the moduli couple to each other through the GVW
superpotential.  Given that the periods of a Calabi-Yau are
generically quite complicated functions of the potential, one
generically gets a potential in which the complex structure moduli
couple to each other, and the extrema of $W$ are broadly
distributed.  This corresponds to a relatively
unpeaked dependence of $\det |D^2 W|$ on $X$.

However, one could study models where this matrix has a highly peaked
dependence, and these would provide distributions which defy some of
our IIB flux vacua intuition.  In some highly decoupled models, for
example, one can
find masses and couplings which instead have very narrow distributions
highly peaked at the ensemble average.  The authors
of \cite{Arkani-Hamed:2005yv}
proposed a set of conditions in an effective field theory example which
would give such narrow distributions.

This model is based on the assumption the
low-energy scalars of the theory are basically decoupled
from each other.  For example, the potential of this theory
might be written as
\bea
V(\phi) = \sum_i ^N V_i (\phi_i).
\eea
For simplicity, we assume that all potentials $V_i$ are
of the same form and are minimized at only two critical points
$\phi_i ^{\pm}$, such that $V_i (\phi_i ^{\pm})=V_i ^{\pm} =
\bar{V_i} \pm \Delta V_i$.  Defining $\bar{V} = N\bar{V_i}$, we
see that distribution of $V$ is given by the standard Gaussian
form
\bea
\rho(V) ={2^N \over \sqrt{2\pi N} \Delta V_i }
e^{-{(V-\bar{V})^2 \over 2\Delta V_i 2}}.
\eea
Indeed, the general form of this Gaussian distribution did not
depend on our assumption that all $V_i$ were of the same form,
or that each had only two minima.  The standard arguments of
the Central Limit Theorem tell us that if $N$ is sufficiently
large, then for any choice of the $V_i$'s the distribution of
$V$ will be a Gaussian centered at the average value and with
a width determined by the root-mean-square of the spread between
the minima of each $V_i$.  The only requirement for this
Gaussian distribution is that the individual potentials $V_i$
be independent.  In such a case, for fixed values of
the complex structure moduli $z_i$, one finds that
$\det |D^2 W|$ is highly peaked as a function of $X$.

The point is that, even though this toy model has $2^N$ vacua,
the value of the potential $V$ is not broadly
distributed among them.  Indeed, the distribution
is a Gaussian which is very sharply peaked at $\bar{V}$.

In a similar way, one can find a broad array of physical observables
whose distribution among vacua is sharply peaked, provided they
depend on moduli which decouple from each other.  For example, if
the Higgs field has couplings to the closed string scalars of
the form
\bea
\sum_i ^N F(\phi_i) h^{\dagger} h,
\eea
then one would also expect the Higgs mass to have a distribution
which is a sharply peaked Gaussian.

In this type of ensemble, the distribution appears to be
flat only in a region very close to the average, and the size of
this region ${\delta X \over X_{avg}}$ is only significant if
the parameter $X$ in question has an average value which is zero,
or very small.  There is usually no natural
reason for any parameter to have
an average value of zero unless there exists a symmetry (perhaps
very slightly broken).  Thus we would find that certain parameters
which were protected by symmetries would appear to ``scan" a broad
distribution in the flux vacua, while other parameters would exhibit
a distribution sharply peaked at a non-zero ensemble average value.

It may appear that this type of an ensemble, with closed string
scalar fields which are very weakly coupled, would rarely
occur as a low-energy description of a controllable string
theory flux compactification.  However, it has been
argued that this type of ensemble
may not be so rare after all.  In \cite{Distler:2005hi}, the authors
impose on the low-energy effective action the constraint that the
Planck scale and K\"ahler
metric be radiatively stable, as well as the constraint that the
superpotential is well-described by perturbation theory.  Assuming these
constraints, they find a minimum width for the distribution of $|W|^2$
which is consistent with a sharply peaked distribution.  Of
course, this is only a minimum width for the distribution;  the actual
width could be much broader.  But the fact that the minimum is
quite small and there is no clear way of arguing for a broader width
suggests that there may exist a sizable class of string flux
compactifications for which some low-energy parameters are not broadly
distributed among flux vacua.

The authors of \cite{Arkani-Hamed:2005yv} raise
this possibility as a concrete way of understanding
the Weinberg argument\cite{Weinberg:1987dv} for
solving the cosmological constant problem via
the Anthropic Principle.  The point would be that such narrow distributions
might explain why one can hold parameters of the Standard Model
fixed while only varying the cosmological constant in Weinberg's
argument.  Given our emphasis on statistics
as opposed to selection, we do not focus on this point
but rather on fact that there is a unique phenomenology associated
with this class of flux vacua, whose statistical distributions are
different from what one typically expects of the IIB ensemble.

\subsection{Supersymmetry breaking scale}

One of the questions which we would like to study is the distribution
of supersymmetry breaking scales on the landscape (this has already
been discussed by several authors\cite{highsusybreak,lowsusybreak}).
It is
important to remember that the question is not necessarily which scale of
supersymmetry does string theory ``predict."  This requires a knowledge
of the probability measure on the set of flux vacua, which we do not
yet understand.  Instead, the questions are: what is the distribution of
the supersymmetry breaking scale in different controlled ensembles of vacua,
how does it correlate with other observable properties, and how many viable
string models do we believe exist with various scales of supersymmetry breaking.

The mathematical
technology does not yet exist which permits a definitive answer to the
question of how the SUSY-breaking scale is distributed in the complete
ensemble of phenomenologically viable vacua.  But we can discuss how the
supersymmetry breaking scale is distributed in various controlled
subensembles.

As pointed out by Douglas and Susskind, the ensemble of flux vacua with
tree-level supersymmetry breaking is likely to be dominated by vacua with
high-scale supersymmetry breaking.  Note that this ensemble is quite different
from the original ensemble which we discussed.  In that ensemble, we looked
at Type IIB vacua where our solutions solved $DW=0$, where the superpotential
$W$ included both the tree-level GVW term and certain non-perturbative
corrections.  It was then assumed that supersymmetry could be broken later by
the addition of small SUSY-breaking terms.  Instead, we now consider the case
where $DW \neq 0$ even at the tree-level approximation.  This ensemble is often
referred to as the ``non-supersymmetric" ensemble, even though our previous
``almost supersymmetric" ensemble also
involved broken supersymmetry at the end of the
day.  But remembering our earlier criterion,
the ``almost supersymmetric" ensemble involved
solutions for which there was a limiting procedure in which the supersymmetry of
the deformed solution was restored.  In the non-supersymmetric ensemble, this is
not the case.

It is relatively easy to see why the non-supersymmetric ensemble is
dominated by high-scale supersymmetry breaking.  In general there can be
many different $F$ and $D$ terms which contribute to supersymmetry breaking,
and we find the general formula
\bea
M_{susy}^4 \sim \sum_i ^{N_F} |F_i|^2 +\sum_j ^{N_D} D_j^2.
\eea
One might imagine writing the number of non-supersymmetric vacua with supersymmetry
breaking scale $M_{susy}$ as\cite{highsusybreak}
\bea
d{\cal N}_{M_{susy}} &=& dM_{susy}^4 \, \int d^{2N_F}F \, \int d^{N_D}D
\, \delta(M_{susy}^4-\sum_i ^{N_F} |F_i|^2 -\sum_j ^{N_D} D_j^2)
\, \rho (F,D)
\nonumber\\
&\sim & (M_{susy})^{2N_F +N_D -1} dM_{susy}^2
\eea
by assuming at the end that the vacuum
density $\rho$ is relatively constant.
Such an answer would indicate that the number of vacua can with
supersymmetry breaking at the scale $M_{susy}$ can grow as a high
power of $M_{susy}$.  Unfortunately this derivation, though intuitively
appealing, is not quite correct\cite{highsusyexcept}.
The basic problem is that $\rho (F,D)$ is
not relatively constant.

To understand this difficulty, we must remember that there are several
further constraints which we must impose to ensure that we are counting
valid solutions.  In particular, we must demand $V' =0$, metastability
(the absence of tachyons in the expansion about the extremum), and
$\Lambda \ll |F|^2 \ll M_{pl} ^4$ (i.e., weak supersymmetry breaking and
very small cosmological constant).

The constraint $V'=0$ implies that all
of the $F$-terms cannot be set independently.  In fact, the $F$-terms
generically fill
out only a one-complex dimensional space, as opposed to the naive
$n_F$-complex dimensional space.  Essentially, the constraint that the goldstino
(which is eaten by the gravitino) have a definite mass
requires that $\langle F_i \rangle$ (which must be in the same multiplet) have
only one complex degree of freedom.  If we write the superpotential as a function
of this multiplet $\Phi$ (without loss of generality, we may choose this multiplet
so that the minimum of the potential occurs when the scalar component vanishes),
we find
\bea
W &=& W_0 + \alpha \Phi + {\beta\over M_{pl}} \Phi^2
+{\gamma \over M_{pl}^2}\Phi^3 +....
\eea
Our above constraints then imply\cite{highsusyexcept}
$\alpha ,\beta ,\gamma \sim |F| \sim M_{susy}^2$
(we ignore $D$-term contributions here for simplicity).
The constraint of very small cosmological constant further implies that
$W_0 \sim |F| \sim \alpha ,\beta ,\gamma$.
We have already seen that $W_0$ appears to have a uniform distribution in many
examples (in the limit of small $W_0$)\cite{Denef:2004ze}.
Having imposed the
necessary constraints, it thus appears that the parameters $\alpha$, $\beta$,
and $\gamma$ also have distributions which are relatively uniform.
We can now write an estimate for the
number of vacua with small cosmological constant as
\bea
d{\cal N}_{M_{susy}} (\Lambda \leq \Lambda_0)
&\sim& {dM_{susy}^2 \over M_{susy}^2}\,
\int d^2 \alpha \, d^2 \beta \, d^2 \gamma d^2 W_0
\theta(\Lambda_0 -V) \theta(\{|\alpha|,|\beta|,|\gamma| \}-M_{susy}^2)^3 \nonumber\\
&\sim& \Lambda_0 M_{susy}^{10} dM_{susy}^2.
\eea
One might expect that the fine-tuning necessary to obtain the
correct Higgs mass is $\sim {M_{higgs} ^2 \over M_{susy}^2}$.
We thus find that the
number of non-supersymmetric vacua with Higgs mass $M_{Higgs}$ and
cosmological constant $\Lambda_0$ generically scales with $M_{susy}$
as $M_{susy}^{10}$.

The story is different, however, in the ``almost
supersymmetric" ensemble.
The reason is essentially because the $F$-terms in this ensemble
are naturally small.  If we imagine them arising from gaugino
condensation (for example), they would damp exponentially as
$M_{susy}^2 =M_{pl}^2 e^{-{const\over g^2}}$.  The
cosmological constant is
given by
\bea
\Lambda = M_{susy}^4 -{3\over M_{pl}^2} |W_0|^2.
\eea
The fraction of states with cosmological constant
$\Lambda < \Lambda_0$ is then given by the
expression\cite{lowsusybreak}
\bea
F_1 (\Lambda < \Lambda_0) &=& \int_0 ^{W_{max}} d^2 W_0 \,
P_W (W_0) \, \int_{\ln(3|W_0|^2)} ^{\ln(3|W_0|^2+\Lambda)}
d(g^2) {1\over g^4} P_{g^2} (g^2)
\nonumber\\
&\sim& \int_0 ^{W_{max}} d^2 W_0 \, {\Lambda_0 \over |W_0|^2}
P_W (W_0) {1\over \ln W_0 ^2} P_{g^2} (-{1\over \ln (W_0)}),
\eea
where $P_W$ and $P_{g^2}$ are the probability densities for
$W$ and $g^2$ respectively.  If these distributions are
roughly flat, then in the limit of small $\Lambda$ we find
\bea
F_1 (\Lambda < \Lambda_0) \propto \Lambda_0 \ln(M_{susy}^2).
\eea
In this ensemble, the growth of the number of vacua as
a function of the
SUSY-breaking scale (at small $\Lambda$) is already less
steep than the power-law expansion of the non-supersymmetric
ensemble.  But we have not yet included any constraints on
the Higgs mass.  If one restricts to the set of vacua
with Higgs
mass in the expected range (as well as small $\Lambda$), the
fraction of vacua with SUSY-breaking scale larger than the
intermediate scale is very small.  The reason is essentially
the same as the standard naturalness argument.  If supersymmetry
is broken at the scale $M_{susy}$ and is transmitted to the
visible sector by gravitational effects, then the natural scale
for the Higgs mass is
\bea
M_{nat}={M_{susy}^2 \over M_{pl}}.
\eea
If the Higgs mass is actually measured at $M_{higgs}$, then the expected
fraction of tuning required is
\bea
\eta &=& {M_{higgs} ^2 \over M_{nat}^2} =
{M_{higgs} ^2 M_{pl}^2 \over M_{susy}^4}.
\eea
There is no fine-tuning suppression for $M_{susy} \leq M_{int}=
\sqrt{M_{higgs} M_{pl}}$.  However, for much larger $M_{susy}$, the fraction
of tuning goes as ${1\over M_{susy}^4}$.  Because the growth in
the number of ``almost supersymmetric" vacua at small $\Lambda$ is only
logarithmic in $M_{susy}$, it is not enough to overcome the power-law
suppression arising from the necessity of tuning the Higgs.

In the subset of vacua where $R$-symmetry is unbroken, there may
be even further enhancements associated with the phenomenological
condition of a small cosmological constant.  For the ``almost
supersymmetric" ensemble, the vanishing of the tree-level
cosmological constant is equivalent to
the vanishing of the tree-level superpotential, $W_0 =0$.  $R$-symmetry
is a discrete symmetry which would guarantee this (as the superpotential
is charged under $R$-symmetry).  But in the ``almost supersymmetric" ensemble,
supersymmetry can still be broken dynamically at a scale $M_{susy}$, which
may also be associated with dynamical $R$-symmetry breaking.  In this
case one would naturally expect the relations
\bea
|DW|^2 \sim M_{susy}^4 \qquad 3{|W|^2 \over M_p ^2} \sim m_{3/2} ^2
M_p ^2 ,
\eea
where $m_{3/2}$ is the gravitino mass.
If $M_{susy}^2 \sim m_{3/2} M_p$,
then these terms are of the same order.  Note that this is precisely
the case of intermediate scale supersymmetry breaking, which is what
would arise if the dominant transmission mechanism of supersymmetry
breaking to the Standard Model sector is gravity.  In this case, we
find that the fine-tuning required to fix an appropriately small
cosmological constant is only ${\Lambda \over M_{int}^4}$.  It can
never be smaller than this (due to the relationship between $|W|$ and
the gravitino mass).  But for a larger supersymmetry breaking scale,
more fine-tuning (${\Lambda \over M_{susy}^4}$) would be required.

Note of course that in models such as this, the enhancement is greater
for smaller values of the gravitino mass.  As such, there are many
more vacua in this class (tree-level $R$-symmetric) with extremely small
gravitino mass than there are with phenomenologically viable gravitino
mass.  But this should not bother us terribly.  We have already conceded
that our universe is highly non-generic, and we are not
attempting to find a
selection principle which selects an allowed value of the gravitino mass.
The name of the game is to consider gravitino masses which are allowed by
experiment, and see if we can estimate the number of models obtainable by
various constructions which
are consistent with this constraint.  We find that in
constructions with tree-level supersymmetry breaking, most viable models
have high-scale SUSY-breaking.  For models in the ``almost supersymmetric"
ensemble, most viable vacua have
intermediate scale supersymmetry-breaking.  This
fraction is even larger among the ``almost supersymmetric"
ensemble vacua for which
there is a tree-level $R$-symmetry.

One can then ask how large is the suppression required to obtain
vacua with unbroken tree-level $R$-symmetry.  Dine
et al.\cite{Dine:2005gz} have
studied algebraic constructions of Calabi-Yau manifolds, and found
in all cases they studied that discrete symmetries remained unbroken when
at most one-third of fluxes were turned on.  This is quite significant,
because it implies a suppression factor of order $(L_*)^{{2\over 3}n}$,
where $n$ is the number of complex structure moduli (and
$4n+4$ is the overall
number of fluxes).
Dewolfe\cite{DeWolfe:2005gy} further argues for the
suppression of the number of vacua with $W=0$ (required for unbroken
$R$-symmetry) in a large algebraic class, though the ratio of
exponents may be subleading in $n$.  For sufficiently large
$L_*$ and $n$, the
suppression required to obtain tree-level $R$-symmetry outweighs
the enhancement in the fraction of $R$-symmetric vacua with viable
Higgs mass and cosmological constant.

Thus far in the analysis, we have assumed a relatively flat distribution
for the complex structure moduli and axio-dilaton.  In fact, the
distribution of these moduli is determined by the curvature of the
moduli space, and can be concentrated at singularities, such as the
conifold.  As was shown in KKLT, if supersymmetry is broken
by adding anti-D3-branes, then they are naturally pushed to the end
of the warped throat, generating an exponentially small SUSY-breaking term
in the potential.  This biasing of the distribution acts in favor
of providing more vacua with low-energy supersymmetry breaking.

Note that we have restricted ourselves to the class of models which
retain supersymmetry at the Kaluza-Klein scale and the string
scale.  One expects that
there are solutions where supersymmetry is broken even at the KK or
string scale\cite{kkstringsusybreak,Saltman:2004jh}, and there may be
many of them.  It would be interesting to compare the distribution
of the supersymmetry breaking scale in the larger ensemble (where
supersymmetry above these higher scales
is not assumed) to the ensembles
we have already considered (which themselves include branches
where high-scale supersymmetry breaking is dominant).

%***************************
%***************************
%***************************
%***************************

\section{Distributions of Standard Model-like Vacua}

We have already discussed flux vacua distributions under
the phenomenological constraints of small cosmological
constant and reasonable Higgs mass.  The next constraint to
impose would be to demand the appearance of Standard Model
gauge group and matter content.  This constraint has been
studied in the Type IIB
context\cite{IDB,Cvetic:2005bn,Marchesano:2004yq,Marchesano:2004xz,Gmeiner:2005vz}.

In the context of Type IIB string theory, the simplest
way to arrive at the SM gauge group and matter content would
be from the open string sector arising from intersecting
D-branes.
However, these space-filling
D-branes contribute to the RR-tadpole constraints, just as the
orientifold planes do.  Any excess negative
D3-brane charge can be canceled
off by the charge induced from RR and NSNS-fluxes.
But these fluxes also serve to generate
a tree-level GVW superpotential which can fix the complex structure
moduli.  In this way, the intersecting brane world story is joined
to our phenomenological study of the
landscape\cite{Dijkstra:2004cc,Blumenhagen:2004xx}.  We briefly
review this construction here.

Our starting point is again Type IIB string theory compactified on
an orientifolded Calabi-Yau 3-fold.  The simplest and most studied
case is actually an orbifold, $T^6 / Z_2
\times Z_2$.  But these considerations apply more generally.
The orientifold action
leaves us with some set of orientifold planes which contribute
to the RR-tadpoles.  Their contributions must all be canceled as
before by some combination of branes and flux-induced charge.
Generally we have several
RR-tadpole conditions to satisfy.  We can do so with a variety of
branes, and the gauge theory arising from these branes should include
a sector which carries Standard Model gauge group and matter content.

\subsection{Standard Model constructions}

The gauge group and matter content of the branes
depend on the type of orientifold we study and the choice of cycles
on which we wrap the D-branes.  For concreteness and simplicity, we
choose the orientifold of $T^6 / Z_2 \times Z_2$ (this
has also been studied
in \cite{Gopakumar:1996mu,T6Z2Z2}).  We then only have
O3- and O7-planes, as well as 3 K\"ahler moduli and
51 complex structure moduli.  We will add so-called ``magnetized"
D-branes\cite{Cascales:2003zp}, in which the gauge
theory on the D-brane has magnetic fields turned on, inducing
lower brane charge as well.  The orientifold action
ensures that the D5-brane and D9-brane charge of any magnetized brane
cancels against the charge of the orientifold image.  As a result,
we need not worry about the corresponding RR-tadpole
conditions.\footnote{There is a subtlety here.  There is
in fact a remaining K-theoretic
$Z_2$ tadpole constraint, which one must be sure to
satisfy\cite{Uranga:2000xp,Witten:1982fp,Witten:1998cd}.}  But the
total D3-brane and D7-brane charges carried by these magnetized branes
must cancel against the charges of the orientifold planes and fluxes.

The gauge group and matter content of any set of branes is
basically determined by the cycles which
they wrap, and the topological intersection
numbers between these cycles.  We consider the case
where all branes are fixed either
at orbifold or orientifold planes (this ensures that we have
an odd number of generations\cite{IDB}).  If a stack of $N$ branes lies
at an orbifold fixed point, the corresponding gauge group is
$U({N\over 2})$, while if they lie at orientifold fixed plane the gauge
group is $USp(N)$.  In either case, $N$ must be even.

The chiral matter content is determined by the topological intersection
number $I_{ab}$ between the
branes.\footnote{It is often easier to think of these branes
in terms of a mirror picture,
where they appear as D6-branes wrapping special Lagrange 3-cycles,
with no magnetic fields.  $I_{ab}$ is then simply the oriented
intersection number.
When we think of a CY mirror in this context, we are first
setting the fluxes to zero.  This is justified
for our purpose here, as it does not affect the open string theory gauge
group or matter content.}
If two branes $a$ and $b$ intersect, then there
are $I_{ab}$ chiral multiplets transforming in the bifundamental
$(G_a , \bar{G}_b)$.  If $b'$ represents the orientifold image of $b$,
then we also find $I_{ab'}$ chiral multiplets transforming
in the $(G_a , G_b)$.  If a brane intersects its own orientifold image,
we have additional chiral multiplets transforming in the symmetric
and anti-symmetric representation of the appropriate gauge group.

The idea is to find two sets of branes, which we call
''visible sector branes" and ``hidden sector branes."  The visible
sector branes contain the gauge group of the Standard Model
and it's chiral matter content, while the hidden sector branes
ensure that the D7-brane RR tadpole conditions are
satisfied.\footnote{There may
be chiral matter which is charged under both visible
sector brane gauge groups and hidden sector brane gauge groups.  This
matter will be referred to as exotic.}  The
D3-brane RR tadpole condition must be undersaturated, with the
difference being made up by charge induced from the RR and NSNS
3-form fluxes.  As we have seen, the number of flux vacua increases
with the amount of flux which we can turn on.

But as before, we wish to find constructions which are supersymmetric
(so that we can later add in lower scale supersymmetry-breaking by
hand).  So if we add several stacks of branes, we wish to be sure that
they respect the same supersymmetries\cite{Ohta:1997fr}.
This condition is most easily
understood from the the dual IIA picture, in which we have D6-branes
with no magnetic field.  In this case, as shown in \cite{Berkooz:1996km},
the condition for two branes to preserve the
same supersymmetry is that they are related by a rotation of
$0 \,{\rm mod}\, 2\pi$ (this is equivalent to the vanishing of
the NSNS tadpole
constraint).
Note however, that violation of this condition does not show that
supersymmetry is broken; it merely shows that a Fayet-Iliopoulos
term is turned on.  Roughly, we have $|\delta \theta| \sim |\xi_{FI}|$.
For each brane (indexed by $j$), we
have\footnote{There is no summation on $j$.} a $D$-term
potential\cite{Kachru:1999vj}
\bea
V_{D}^j = (\sum_i q_i ^j |\phi_i ^j|^2 -\xi^j)^2 ,
\eea
where the scalars $\phi_i ^j$ arise from open strings stretching
between branes $i$ and $j$, and $q_i ^j$ are the scalar charges under
the $U(1)$ of
brane $j$.
We see that if the NSNS tadpole constraints are satisfied, then
$\xi^j=0$ and the $D$-term potential will vanish if we also set
$\phi_i ^j =0$.  However, even if some of the $\xi$'s are non-zero,
some of the open string scalars $\phi$ will become tachyonic if their
charge $q$ has the appropriate sign.  In this case, the scalars
can get vevs which cancel the FI-term and set the $D$-term
to zero.  This process is known as brane recombination; the veving
of the tachyonic open-string scalar $\phi_i ^j$ corresponds to the
branes $i$ and $j$ forming a bound state.

Bearing in mind that this brane recombination process can occur, we
find that it is not necessary for the NSNS tadpole conditions to be
satisfied in order to maintain supersymmetry.  Indeed, if we have $r$
branes the we have $r$ $D$-terms, but we generically
have $\sim r^2$ open string scalars (since our branes
generically have non-zero
topological intersection with one another).  In this case, we
generically find for any brane a scalar charged with either sign of
charge.  In such a case, we generally can restore supersymmetry
by some type of brane recombination for any values of the FI-terms.

\subsection{Some results}

One particular construction of visible and hidden sector branes was
found by Marchesano and
Shiu\cite{Marchesano:2004yq,Marchesano:2004xz}, and generated a
visible sector yielding an $U(4) \times SU(2)_L \times
SU(2)_R$ extension of the Standard Model.  They
searched for models in which all of the FI-terms vanished,
thus avoiding some of the complications of brane recombination.
Their model is described by the brane embedding
\bea
N_a=6 & (1,0)(3,1)(3,-1) \nonumber\\
N_b=2 & (0,1)(1,0)(0,-1) \nonumber\\
N_c=2 & (0,1)(0,-1)(1,0) \nonumber\\
N_d=2 & (1,0)(3,1)(3,-1) \nonumber\\
N_{h1} =2 & (-2,1)(-3,1)(-4,1) \nonumber\\
N_{h2} =2 & (-2,1)(-4,1)(-3,1) \nonumber\\
N_{h3} =8 & (1,0)(1,0)(1,0)
\label{visible sector}
\eea
where the wrapping numbers $(n,m)$ describe the integer
quantized magnetic flux and winding, respectively,
of the brane on each of the three tori.
The first 4 brane stacks are the visible sector and the
last three brane stacks are the hidden sector.  This model
allows one to obtain one
quantized unit of D3-brane charge from fluxes (which thus
fix the complex structure moduli and axio-dilaton).  Indeed,
it was shown in \cite{Kumar:2005hf} that this is in fact the
only model on this orientifold with the given choice of
visible sector, non-zero
flux and vanishing FI-terms.

However, several choices of hidden sector can be exhibited
with allow much larger
amounts of flux, once brane recombination is utilized (i.e.,
we drop the demand for vanishing FI-terms).  Several of
these models where exhibited\cite{Kumar:2005hf,web page},
the largest of
which exhibited $N=9$ units of quantized D3-brane charge
induced from the fluxes.  The counting arguments
we discussed before suggest that we would
find $\sim 10^{30}$ flux vacua in this model with the chosen
visible and hidden sector brane content.

There are in fact several other choices of visible sector
which produce different extensions of the Standard Model,
and they can be studied in a similar vein\cite{Cvetic:2005bn}.

\subsection{Standard Model phenomenology considerations}

We see that we can potentially find large numbers of flux
vacua with the Standard Model gauge group and chiral matter
content.  One can then study the distribution of parameters
such as the tree-level cosmological constant
for these models as
reviewed earlier.  The hope is to utilize
the large number of vacua to estimate which fraction of these
vacua happen to have low-energy parameters which are
within experimental precision of the real world.  For example,
Douglas\cite{Douglas:2003um} estimates that a fine-tuning
of parameters on the order of $10^{238}$ is need to ``get the numbers
right," i.e., to fine-tune the actual masses and couplings
(plus cosmological constant) from their natural values to
experimentally measured values with current precision.  Thus, if
we find a number of vacua
${\cal N}$ with the right SM matter content such
that ${\cal N} \gg 10^{238}$,
and the distribution of scalar vev's is fairly uniform, then
we may be quite confident that there exists a string vacua whose
low-energy parameters match the real world to within experimental
precision.  We are also in a position to study the phenomenology
of this set of vacua.

An extended discussion of this phenomenology is beyond the
scope of this review\cite{Kumar:2005hf}, but there are
a few points worth noting.  Our initial paradigm was to fix
the moduli at a scale well above the supersymmetry breaking
scale, in order to ensure that non-supersymmetric corrections
to the potential did not destabilize the solution.
But we do not include here the open string scalars which are
charged under the Standard Model.
These scalars must
arise from open string non-vector-like
chiral multiplets, and the scalars
for these multiplets cannot get mass except at the SUSY-breaking
scale.  These scalars are coupled to the K\"ahler
moduli through the $D$-term equations.  If the K\"ahler moduli
are not fixed by $F$-terms, then they might be fixed by
$D$-terms instead, but this would be at the SUSY-breaking
scale.  This could be problematic for the stability of the solution.
Alternatively, if the K\"ahler moduli are fixed by
$F$-terms, then the open string scalars would likely break
supersymmetry by a combination of $F$-terms and $D$-terms.
Interesting features of supersymmetry breaking in this context
are also discussed in \cite{susy breaking,Choi:2004sx}.

Ubiquitous features of these types of constructions are
non-chiral exotics, extra $U(1)$'s and potential discrete
symmetries.  We can study the distribution of these
properties among the class of vacua which admit the Standard
Model.  This counting can thus be convolved with the
usual phenomenological discussion of these features.

%***************************************
%***************************************
%***************************************
%***************************************

\section{Discussion - an Infinity of Vacua}

We have seen that there are many very concrete calculations we
can perform, from which we can derive
the distribution of low-energy
properties among flux vacua in a variety of specific constructions.
The question is, ``how can we use these distributions to make
contact with phenomenology?"  Unfortunately, our limited understanding
of non-supersymmetric string compactifications makes it difficult for
us to make progress.  But we have already seen that the study of
these distributions does provide us clues which can direct further
phenomenological work.  In addition, we can also study some
formal string theory questions which have significant
implications for the role of string theory in phenomenology.

One of the questions we hope to better answer with this study
of the landscape is how many string vacua match the real
world to experimental precision, and in particular, is
the number finite or infinite\cite{Douglas:2005hq}.  One
may be tempted already
to speculate on the implications of the answer to this
question\cite{Douglas:2003um}.

It is sometimes treated as disastrous for the predictivity of
string theory if the number of phenomenologically viable vacua
is infinite.  However, that is not necessarily the case.
Even if the number of viable vacua is infinite, it can still
be possible to make predictions
regarding the values of masses and low-energy
couplings (in principle).  Consider the parameter space
of a low-energy description of string theory vacua.  This can
be thought of as the parameter space of the low-energy effective
field theory, and is consequently infinite dimensional (as the
number of possible couplings in effective field theory is
infinite).  The real world, to within experimental uncertainty,
occupies some ball\footnote{By ``ball," we simply mean an open
set.  It can of course have a highly complicated shape,
depending on how the space is parameterized.}
in this space.  As experimental data
becomes more precise, the size of the ball shrinks.
The various consistent vacua of string theory appear as
points (not necessarily isolated) in this space.  If
the ball shrinks to a size such that there is exactly one string
vacuum inside the ball, then string theory gives a unique prediction
for which vacuum we live in and what its low-energy properties are
(to arbitrary accuracy).  Even if there are an
infinite number of isolated flux vacua, this will always
occur in principle
as long as the set of string vacua do not have limit points
in the ball.

We may have a situation in which, for any open set in parameter space
containing
the theory which exactly describes the real world, there exists at least
one string vacuum within that open set.
Even in this case, it is still possible in principle for string theory
to make useful predictions of low-energy couplings.  For
example, suppose we are only interested in two parameters: the
electron mass and the proton mass.  It may be that the infinite
number of phenomenologically viable vacua are effectively confined to
a 1-dimensional surface in this two-dimensional parameter subspace.  This
means that we would not be able to predict either the electron mass
or the proton mass individually.  But if new experiments gave increased
precision in the measurement of the electron mass, this would translate
into a prediction for the next order of precision in the proton mass.

Of course, it is also possible that the set of vacua
densely fills the space of low-energy
parameters.\footnote{For some contrary arguments, see \cite{notdense}.}
In this case, it is true that string theory is unable to predict
any low-energy parameters, or find any correlations between them.  It
is interesting to note the implications of this result for the Bekenstein
bound.  The Bekenstein bound\cite{Bekenstein:1980jp,Bekenstein:1974ax}
on the entropy of a system is given by
\bea
S \leq {2\pi ER \over \hbar c}.
\eea
One may show that this bound can be trivially violated by
field-theoretic systems
(for example, a very large number of scalars very
weakly coupled to gravity\cite{Page:1982fj}).  But one may
be suspicious of these
counter-examples, because there is no known realization of them in a
consistent theory of quantum gravity, such as string theory.  If it
were in fact true, however, that the set of string vacua densely fills
the space of all possible low-energy effective field theories, then
these counter-examples would be legitimate and the Bekenstein bound
would fail.  One may view this as a way for the string landscape to
provide evidence for or against the Bekenstein bound.  Alternatively,
one may view the circumstantial evidence in favor of the Bekenstein
bound as evidence for the claim that string vacua do not densely pack
the space of possible low-energy effective field theories.

This question addresses the formal predictivity of string theory.  But
a related, and arguably more relevant question is whether experimental
precision can ever give us constraints tight enough (that is, a small
enough ``ball") to allow string
theory to make useful predictions.  Alternatively, one might ask if
string theory could exclude, as a practical matter,
any features which model-builders find
interesting.  It may be easier for string theory to provide input to
model-building by determining that particular classes of constructions
are unlikely (or quite likely) to yield vacua exhibiting particular
interesting features.  Current studies of distributions on the landscape
already give hints as to how this program can proceed.

These are just a few examples of the variety of questions, both formal and
phenomenological, which we can study via the landscape.  Clearly,
there is much more work which can be done is this emerging field.
The most exciting results and prospects appear to lie in the future.

%%%%%%%%%%%%%%%%%%%%%%%%%%%
\section*{Acknowledgements}
We gratefully acknowledge T. Banks, F. Denef, M. Dine,
M. Douglas, S. Kachru, F. Larsen, S. Sethi,
G. Shiu and L. Susskind for useful discussions.  We are especially
grateful to J. Wells for discussions and collaboration at all stages
of our study of this field.  This work is supported in part by
NSF grant PHY-0314712.

\end{document}